\documentclass[aps,preprint,superscriptaddress,showpacs,showkeys,preprintnumbers,amsmath,amssymb]{revtex4}
\usepackage[dvips]{graphicx}
\usepackage{dcolumn}
\usepackage{amsmath}
\usepackage{amssymb}
\usepackage{amsbsy}
\usepackage{units}

\usepackage{bm}

\usepackage{color}
\usepackage{ulem}

\begin{document}
\title{Ferromagnetism and impurity band in a new magnetic semiconductor: InMnP}
\author{M. Khalid}
\affiliation{%
Helmholtz-Zentrum Dresden Rossendorf, Institute of Ion Beam Physics and Materials Research, Bautzner Landstrasse 400, D-01328 Dresden, Germany}%
\author{Eugen Weschke}
\affiliation{%
Helmholtz-Zentrum Berlin f\"{u}r Materialien und Energie, Wilhelm-Conrad-R\"{o}ntgen-Campus BESSY II, D-12489 Berlin, Germany}%
\author{W. Skorupa}
\affiliation{%
Helmholtz-Zentrum Dresden Rossendorf, Institute of Ion Beam Physics and Materials Research, Bautzner Landstrasse 400, D-01328 Dresden, Germany}%
\author{M. Helm}
\affiliation{%
Helmholtz-Zentrum Dresden Rossendorf, Institute of Ion Beam Physics and Materials Research, Bautzner Landstrasse 400, D-01328 Dresden, Germany}%
\affiliation{%
Technische Universit\"{a}t Dresden, 01062 Dresden, Germany}%
\author{Shengqiang Zhou}
\affiliation{%
Helmholtz-Zentrum Dresden Rossendorf, Institute of Ion Beam Physics and Materials Research, Bautzner Landstrasse 400, D-01328 Dresden, Germany}%
\date{\today}
\begin{abstract}
We have synthesized ferromagnetic InMnP, a member of III-Mn-V ferromagnetic semiconductor family, by Mn ion implantation and pulsed laser annealing. Clear ferromagnetic hysteresis loops and a perpendicular magnetic anisotropy are observed up to a Curie temperature of 42 K. Large values of negative magnetoresistance and magnetic circular dichroism as well as anomalous Hall effect are further evidences of a ferromagnetic order in InMnP. An effort is made to understand the transport mechanism in InMnP using the theoretical models. We find that the valence band of InP does not merge with the impurity band of the heavily doped ferromagnetic InMnP. Our results suggest that impurity band conduction is a characteristic of Mn-doped III-V semiconductors which have deep Mn-acceptor levels.
\end{abstract}

\pacs{75.50.Pp,75.70.-i} \keywords{Ferromagnetism in InMnP} 
\maketitle

Diluted magnetic semiconductors (DMS) containing a few atomic percent of a magnetic element exibit simultaneously magnetic and semiconducting properties below the Curie temperature. The III-V based prototype ferromagnetic semiconductor GaMnAs has revealed a variety of unique features induced by the combination of its magnetic and semiconducting properties. Among these are the electric-field control of the Curie temperature \cite{Ohno944} and the magnetization direction \cite{Chiba515} and the low current density induced magnetization reversal \cite{Watanabe082506}, which are expected to be useful for upcoming ultralow-power spintronic devices.\\
Yet a controversy still remains as to the GaMnAs band structure and the mechanism behind the ferromagnetism. Generally two models are being used to explain the ferromagnetism in III-V semiconductors: The mean-field Zener model, where the ferromagnetism in III-V:Mn arises due to the $p-d$ exchange interaction between the valence band (VB) holes and the localized Mn-3$d$ electrons, and the Zener double-exchange model, where the hopping of the spin-polarized holes in the impurity band (IB) stabilizes the ferromagnetism. The former model seems to explain several features of GaMnAs such as carrier dependent metal-insulator transition \cite{Macdonald195}, hole concentration dependent Curie temperature \cite{Nishitani045208} etc. On the other hand, optical \cite{Hirakawa193312,Burch087208,Ando067204} and transport \cite{Rokhinson161201} studies of GaMnAs have revealed that the Fermi energy level (E$_f$) lies in the IB within the bandgap of GaMnAs. Despite massive investigations on the origin and control of ferromagnetism in DMS, the controversy on the origin of ferromagnetism and the location of impurity-band in III-V-Mn semiconductors still remains \cite{Ohya342,Burch087208,Masek227202,Smarth360,Dobrowolska444}. Note that the Mn energy level ranges from 10 meV to 440 meV in different III-V hosts. It is unlikely that all different III-V-Mn combinations can be treated within a single model.\\
Scarpulla et~al. prepared ferromagnetic GaMnP which has the deepest Mn acceptor level (440 meV) among III-V semiconductors. They found that the ferromagnetic exchange coupling was mediated by holes localized in a Mn-derived band that is detached from the valence band \cite{Scarpulla207204}. Due to the deep Mn-acceptor level, strongly localized hole states are expected, leading to a detached impurity band in GaMnP. Moreover, GaMnP has an indirect bandgap which also restricts its applications in optoelectronic devices. Therefore, it is worthy to choose a system that has a bandgap and Mn-acceptor energy level in between those of GaMnP (440 meV) and GaMnAs (110 meV) for a further understanding of the valence-impurity band dilemma in III-V semiconductors.\\
InP is a suitable candidate for investigating the existence and merging of the IB in InMnP, because it has a bandgap (1.34 eV) closer to that of GaAs (1.42 eV) and a Mn-acceptor level at an energy 220 meV above the valence band which is alomst half of the GaMnP \cite{Clerjaud3615}. Moreover, Dietl et al. have demonstrated theoretically that Mn-doped InP could be a promising diluted magnetic semiconductor (DMS) with a Curie temperature of $\sim$ 50 K \cite{Dietl1019} in case of 5$\%$ Mn substitute the Indium in crystal grate nodes and of 10$^{20}$/cm$^3$ hole concentration. To the best of our knowledge, there is hardly a comprehensive report available on the magnetic properties of InMnP in literature except some reports which, in our opinion, do not show a clear evidence of carrier mediated ferromagnetism in InMnP. First, undesirable secondary phases appear in samples annealed by conventional methods (e.g. thermal annealing) \cite{Shon1292,Shon1065,Bucsa073912}. Second, ferromagnetism is reported in Zn doped p-type InMnP for a Mn concentration of only 1 at.$\%$ \cite{Shon220}. This would be incompatible with the well-established theory of ferromagnetism in semiconductors \cite{Dietl1019,Dietl377}. There must be a sufficient amount of Mn ions ($\sim$ 5 at.$\%$) in the sample so that a net ferromagnetic order can be achieved at a reasonable high temperature ($\sim$ 50 K). Hence, those previous reports do neither prove the intrinsic mechanism of ferromagnetism nor do they give any insight into the nature of the impurity band in InMnP. Therefore, it is highly desirable to synthesize a diluted magnetic semiconductor, InMnP, with reliable magnetic and transport properties. It can also contribute to the understanding of the valence-impurity band picture in III-V semiconductors.\\
In this letter we focus on the Mn-induced ferromagnetism and the location of Mn-induced impurity band in InMnP prepared by Mn ion implantation and pulsed laser melting. We mainly discuss the results of a heavily doped (Mn 5 at.$\%$) InMnP sample, however, some results of a low Mn-doped (Mn $\sim$ 2.5 at.$\%$) InMnP sample are also included. The magnetization, magnetotransport and x-ray magnetic circular dichroism (XMCD) measurements were carried out to support an intrinsic origin of ferromagnetism and the existence of a separate impurity band in InMnP.\\
Semi-insulating InP(100) wafers were implanted with 50 keV Mn$^+$ to fluences of 2$\times$10$^{16}$/cm$^2$ and 1$\times$10$^{16}$/cm$^2$, hereafter referred as sample A and sample B, respectively. After that the samples were annealed by a XeCl excimer laser using an energy density of 0.4 J/cm$^2$ for a single pulse duration (30 ns). The Mn-concentration over a depth of around 90 nm (measured by transmission electron microscopy and Auger electron spectroscopy, not shown) for sample A was $\sim$ 5 at.$\%$.  Magnetization of the samples was measured by a superconducting quantum interference device (SQUID). Magnetotransport and XMCD measurements were performed using a Lakeshore system and at the beamline UE46/PGM-1 at BESSY II (Helmholtz-Zentrum Berlin), respectively.\\   
\begin{figure}[ht]
\begin{center}
\includegraphics[width=0.4\textwidth]{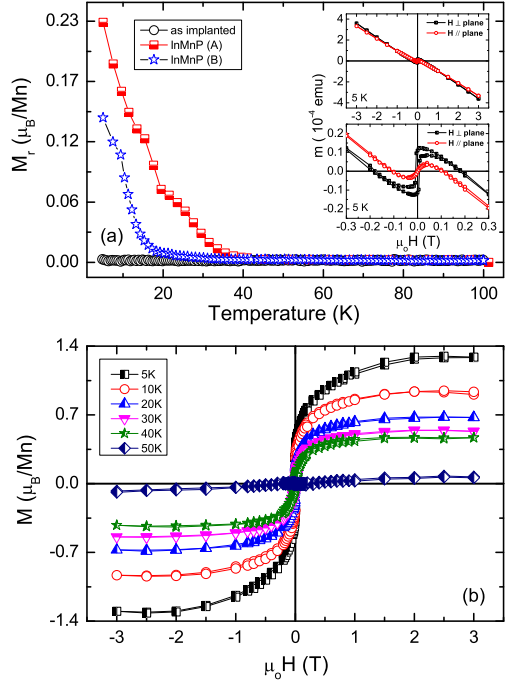}
\caption{\label{Fig.1} (a) Remanent magnetization versus temperature for an as-implanted sample and two laser annealed InMnP samples A and B. Inset shows the magnetization of a laser annealed sample A in magnetic field ranges $\pm$3 and $\pm$0.3 T applied parallel and perpendicular to the sample plane. (b) Mangetization loops of laser annealed sample A measured at several temperatures. The magnetic field is applied perpendicular to the sample plane.}
\end{center}
\end{figure}
Figure~\ref{Fig.1}(a) shows the remanent magnetization versus temperature for three different InMnP samples. The Curie temperature (T$_c$) of the laser annealed samples A and B are found to be $\sim$ 42 K and $\sim$ 20 K, respectively, while the as-implanted sample did not exhibit a ferromagnetically ordered state down to the lowest measured temperature. The inset of Fig.~\ref{Fig.1}(a) represents the magnetic response of sample A as a function of a magnetic field at 5 K. There are two main contributions in the magnetic signals, a diamagnetic response which is due the InP substrate and a ferromagnetic one which originates from the Mn implanted layer. It also shows a magnetic anisotropy (easy axis perpendicular to the sample plane) which often originates due to tensile strain, not only in Mn-implanted samples, but also in III-V semiconductors grown by molecular beam epitaxy (MBE) \cite{Zhou093007,Rushforth073908}. The dependence of Curie temperature on the Mn-concentration supports the carrier mediated ferromagnetism in InMnP.\\ 
Ferromagnetic loops are observed up to the Curie temperature for sample A as shown in Fig.~\ref{Fig.1}(b). The saturation magnetization at 5 K is $\sim$ 1.2 $\mu_B$/Mn by considering that all implanted Mn ions contribute to the magnetization of the sample. Practically, this value is largely underestimated due to the following reasons: First, the sputtering effect in III-V semiconductors during ion implantation reduces the effective implantation fluence \cite{Fritzsche39}. Secondly, it is known that the pulsed laser annealing results in a magnetically inert layer with a large Mn concentration due to surface segregation \cite{Scarpulla073913}. Moreover, self compensation like in GaMnAs \cite{Bouzerar125207} can further depress the ferromagnetism in InMnP. If half of the implanted Mn concentration contributes to the total magnetization of InMnP then the estimated magnetization will be $\sim$ 2.4 $\mu_B$/Mn which is comparable to the magnetization of InMnAs \cite{Zhou093007,Schallenberg042507}.\\
X-ray absorption spectroscopy (XAS) is widely used to probe the electronic structure of matter while x-ray magnetic circular dichroism (XMCD) by utilizing the circularly polarized light gives selective elemental magnetic information of a specimen. We have performed XMCD measurements on sample A at the Mn L$_{2,3}$-edges in which the 2p core electrons are excited to the unoccupied 3$d$ states and consequently information on the electronic structure of polarized Mn band is obtained, for more detail see the article and references therein \cite{Schutz737}. The sample was etched in a diluted HCl solution (5 \%) prior to XMCD measurements to remove the surface oxide layer. This method has also been used previously for GaMnAs and GaMnP \cite{Edmonds4065,Stone012504}.
\begin{figure}[ht]
\begin{center}
\includegraphics[width=0.4\textwidth]{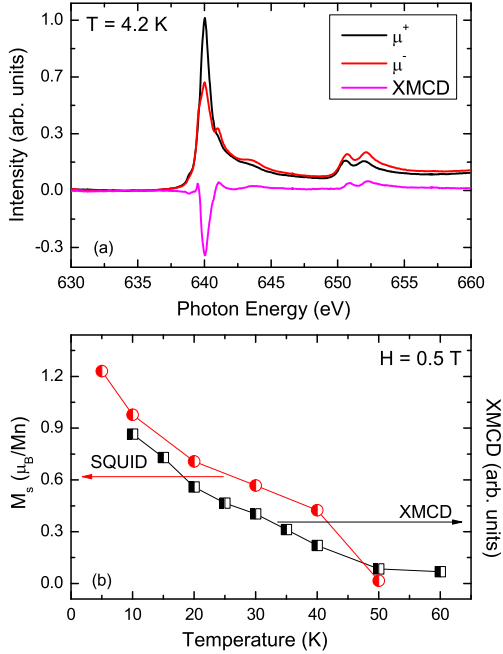}
\caption{\label{Fig.2} (a) XAS and XMCD spectra of a InMnP sample A measured at 4.2 K under a magnetic field of 4 T applied normal to the sample surface. (b) Saturation magnetization and normalized XMCD signals of the same sample at different temperatures under a magnetic field of 0.5 T.}
\end{center}
\end{figure}
\\
Figure~\ref{Fig.2}(a) shows XAS spectra at the Mn L$_{2,3}$-edges along with a XMCD spectrum in total electron yield (TEY) mode at 4.2 K under a magnetic field of 4 T for sample A. The main intense features in the XAS spectra at the L$_{2,3}$-edges (at energies around 640 eV and 652 eV) are due to electric-dipole allowed transitions (i.e. $\Delta\ell$=$\pm1$) from 2$p$ to unfilled 3$d$ states. A careful inspection of XAS spectra of InMnP indicates that the spectral features are very similar to those of a ferromagnetic GaMnP, GaMnAs and InMnAs \cite{Edmonds4065,Stone012504,Scarpulla073913} regardless they have different energy gaps and Mn-acceptor levels. It also hints to a similar bonding and $p-d$ exchange interaction in most of the Mn-doped III-V semiconductors. The XMCD signal at L$_3$-edge of InMnP defined as ($\mu^+$-$\mu^-$)/($\mu^+$+$\mu^-$) is $\sim$ $40\%$ at 4.2 K under a magnetic field of 4 T which is comparable to that of ferromagnetic GaMnP \cite{Scarpulla073913}. The large XMCD signal of InMnP also indicates that it has a high spin polarization at the Fermi energy (E$_f$). The XMCD sum rules also provide information on degree of the spin and orbital moments in the system \cite{Thole1943,Obrien12672}. The spin moment calculated using XMCD data and the sum rules in InMnP sample A is $\sim$ 1 $\mu_B$/Mn while the orbital moment is negligibly small. This value of spin moment is slightly smaller than that obtained from SQUID measurements. The discrepancy in the magnetic moment could be due to the depth profile of the Mn concentration and the depth sensitivity of two techniques (SQUID is volume sensitive while XMCD in the total electron yield mode is surface sensitive).\\
We have compared the magnetization measured by SQUID magnetometry and temperature dependent normalized XMCD signals in order to exlude any contribution from external impurities to the magnetization of the InMnP sample. Figure~\ref{Fig.2}(b) shows the saturation magnetization and normalized XMCD signals at different temperatures.
Both, the magnetization and normalized XMCD signals decrease with temperature and vanish at $\sim$ 50 K just above the Curie temperatue (42 K). The very analogous temperature dependence of both quantities indicates an intrinsic origin of ferromagnetism in InMnP.
\begin{figure}[ht]
\begin{center}
\includegraphics[width=0.4\textwidth]{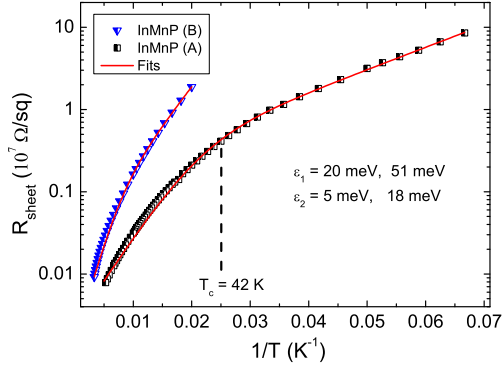}
\caption{\label{Fig.3} Sheet resistance of InMnP samples A and B as a function of inverse temperature. The red solid lines show the fits of data to Eq.~\ref{1}.}
\end{center}
\end{figure}
\\
Another method for studying the carrier-mediated nature of ferromagnetism is the magnetotransport. We have carried out temperature dependent resistivity and magnetotransport (magnetoresistance, Hall effect) measurements on InMnP samples using van der Pauw geometery under a magnetic field perpendicular to the sample plane using a Lakeshore system. The temperature and magnetic field were varied up to 150 K and 6 T, respectively. Figure~\ref{Fig.3} shows the sheet resistance of InMnP samples A and B as a function of inverse temperature at zero field. We did not observed any metal-insulator transition, as has been observed in heavily doped GaMnAs \cite{MatsukuraR2037}, the InMnP samples rather show insulating character similar to that of GaMnP \cite{Scarpulla207204}. Therefore, to describe a thermally activated conduction process at high and low temperatures, we used a model given in Eq.~\ref{1},
\begin{equation}
\label{1}
\rho(T)^{-1}=\left\{\rho_1\exp(E_1/\kappa_BT)\right\}^{-1}+\left\{\rho_2\exp(E_2/\kappa_BT)\right\}^{-1}
\end{equation}
where pre-exponential constants $\rho_1$, $\rho_2$ and activation energies $E_1$, $E_2$ are the free parameters. The measured temperature dependent resistivity data fits to this model quite nicely. The fitting results indicate two thermally activated contributions to the resistivity of InMnP. For sample A a high temperature activation region with an activation energy $\sim$ 20 meV ($E_1$) and a low temperature activation region with an energy $\sim$ 5.1 meV ($E_2$) are observed. The activation energies in InMnP sample A for two activated regions are smaller than that of GaMnP \cite{Scarpulla207204} which are expected because of the following reason. The transition between the valence-band and the acceptor states are responsible for the low resistivity at high temperatures. As the Mn-acceptor level in InMnP lies at lower energy (220 meV) compared to GaMnP (400 meV), therefore, to excite an electron from valence-band to acceptor states needs less energy in InMnP. A close examination of the resistivity data of sample A in Fig.~\ref{Fig.3} also indicates a continuous change in the slope near T$_c$ (42 K) which can be due to the nearest-neighbor hopping transport \cite{Kaminski235210} at low temperatures with an activation energy 5.1 meV in sample A. On the other hand, for the low Mn-doped InMnP sample B $E_1$ and $E_2$ increase to $\sim$ 50 meV and $\sim$ 18 meV, respectively. This is expected due to the narrowing of the impurity band and the shift of the Fermi energy in the Mn-induced band accompanied by a reduction in hole concentration at lower Mn concentration. It seems that, as least for semiconductors with deep Mn-acceptor levels, an impurity band is formed in heavily Mn-doped III-V semiconductors which is detached from the valence band. The high insulating character and the hopping transport in InMnP support a detached impurity band in InMnP.\\
To further shed light on the hopping transport mechanism in InMnP at low temperatures, we have perfomed magnetotransport experiments over a wider temperature range. Figure~\ref{Fig.4}(a) shows the magnetoresistance of sample A measured at different temperatures above and below the Curie temperature. It exhibits negative magnetoresistance as is typical for III-V diluted magnetic semiconductors. At the lowest measured temperature (15 K) the relative magnetoresistance defined as $MR(\%)=\left[\left\{\rho(H)-\rho(0)\right\}/\rho(0)\right]\times100$, is about 64$\%$ at 6 T which increases with the temperature and reaches a maximum value of 66$\%$ at 20 K and then decreaes with the temperature. The negative magnetoresistance in InMnP can be explained as follows: when a magnetic field is applied, it results in an antiferromagnetic coupling between the Mn ion and the hole. Consequently, the wavefunctions of Mn-hole complexes expand and the increased overlapping of these wavefunctions results in a negative magnetoresistance of InMnP. We have used a model in high-field limits i.e. $\lambda$$\ll$$a$, where $\lambda$ and $a$ are the magnetic length and localization radius, respectively, which has been used for insulating GaMnAs previously \cite{Esch13103}. The model is given in Eq.~\ref{2}.
\begin{figure}[ht]
\begin{center}
\includegraphics[width=0.4\textwidth]{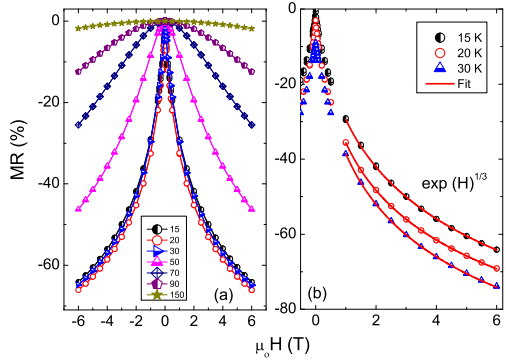}
\caption{\label{Fig.4} (a) Magnetoresistance of a InMnP sample A measured at several temperatures. (b) The red solid lines show the fits of the data to a model given in Eq.~\ref{2}. Note that for clarity, the MR curves measured at 20 and 30 K are shifted vertically by constants 3 and 9, respectively.}
\end{center}
\end{figure}
\begin{equation}
\label{2}
\rho(H)=\rho_o\exp\left[\frac{C}{(\lambda^2T)^{1/3}}\right]
\end{equation}

where $\lambda=\left(\frac{\hbar}{e\mu_0 H}\right)^{1/2}$ is the magnetic length. This model is used to fit the magnetoresistance data below T$_c$ and the results are shown in Fig.~\ref{Fig.4}(b). The fit results show that the model is applicable to explain the magnetoresistance of InMnP in the high-field range of 1-6 T. The fitting parameter $C$ has a negative sign due to the expansion of the wave functions in a magnetic field. The magnetotransport results suggest that hopping is the main conduction mechanism in InMnP at low temperatures which is also an indication of a separated Mn-impurity induced band within the bandgap of InP.\\
Finally, the anomalous Hall effect was measured in the InMnP sample. It reflects the presence of spin polarized carriers along with their type and concentration in the sample. The Hall resistance R$_{Hall}$ of a magnetic semiconductor can be expressed phenomenologically as in Eq.~\ref{3}.
\begin{equation}
\label{3}
R_{Hall}=R_oB+R_sM
\end{equation}

\begin{figure}[ht]
\begin{center}
\includegraphics[width=0.4\textwidth]{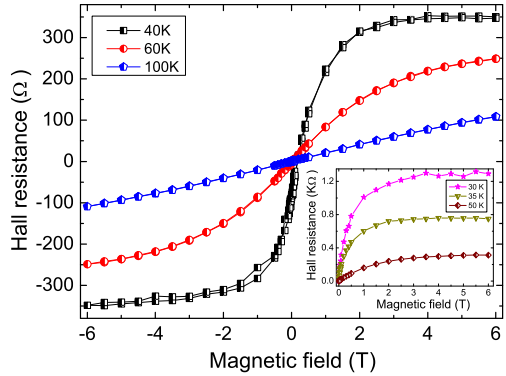}
\caption{\label{Fig.5} Anomalous Hall effect in InMnP at different temperatures. Inset shows the anomalous Hall effect below T$_c$.}
\end{center}
\end{figure}
The first term on the right-hand side of Eq.~\ref{3} is the ordinary Hall resistance while the second term stands for the anomalous Hall resistance. In general the anomalous Hall contribution is proportional to the macroscopic magnetization of the sample \cite{Hurd3}. Figure~\ref{Fig.5} shows the Hall resistance of InMnP sample A as a function of applied field measured at different temperatures. The Hall resistance of InMnP is dominated by the anomalous Hall component at low temperatures, see Fig.~\ref{Fig.5} and its inset. The temperature dependence and the shape of the Hall resistance is quite similar to that of magnetization of InMnP sample A. Therefore, we can conlude that both type of measurements reflect the intrinsic ferromagnetism of InMnP. The slope of the Hall resistance remains positive up to the measured temperatures and fields. This confirms that holes are the majority carriers in InMnP.\\
In conclusion, we have prepared the diluted magnetic semiconductor InMnP by Mn implantation and pulsed laser annealing. The transport results suggest that a separate impurity band formation is preferable even in heavily doped InMnP. The magnetization, x-ray circular dichroism, magnetoresistance and anomalous Hall effect reflect an intrinsic ferromagnetic order in Mn-implanted InP. A large XMCD signal and negative magnetoresisance indicate that InMnP has a high spin polarization at the Fermi energy (E$_F$) and the ferromagnetism is carrier mediated. Last but not least, a careful inspection of XAS spectra of InMnP points that the spectral features are very similar to those of ferromagnetic GaMnP and GaMnAs \cite{Scarpulla207204} regardless they have different energy gaps and Mn-acceptor levels. The high insulating character of InMnP is very similar to that of GaMnP which could be due to strong localization of carriers in an impurity band (low mobility) and hopping transport mechanism support a separate impurity band formation in InMnP and similar III-V semiconductors which have deep Mn-acceptor levels.\\

The authors thank Dr. Ren$\acute{e}$ H\"{u}bner for TEM measurements. The work is financially supported by the Helmholtz-Gemeinschaft Deutscher Forschungszentren (VH-NG-713).

\end{document}